\batchmode
\makeatletter
\def\input@path{{C:/Dropbox/SubmittedPapers/0_OptimalMitigation//}}
\makeatother
\documentclass[11pt,english]{article}
\usepackage{lmodern}
\usepackage[T1]{fontenc}
\usepackage[latin9]{inputenc}
\usepackage{geometry}
\usepackage{babel}
\usepackage{graphicx}
\usepackage{setspace}
\usepackage{esint}
\usepackage[authoryear]{natbib}
\usepackage{subscript}
\usepackage[unicode=true,pdfusetitle,
 bookmarks=true,bookmarksnumbered=false,bookmarksopen=false,
 breaklinks=false,pdfborder={0 0 1},backref=false,colorlinks=false]
 {hyperref}

\makeatletter

\providecommand{\tabularnewline}{\\}

\newcommand{\lyxaddress}[1]{
\par {\raggedright #1
\vspace{1.4em}
\noindent\par}
}

\@ifundefined{date}{}{\date{}}
\makeatother

\begin{document}

\title{Economics of limiting cumulative CO\textsubscript{2} emissions}

\author{Ashwin K Seshadri}

\maketitle

\lyxaddress{Divecha Centre for Climate Change, Indian Institute of Science, Bangalore
560012, India. email: ashwin@fastmail.fm}
\begin{abstract}
Global warming from carbon dioxide (CO\textsubscript{2}) is known
to depend on cumulative CO\textsubscript{2} emissions. We introduce
a model of global expenditures on limiting cumulative CO\textsubscript{2}
emissions, taking into account effects of decarbonization and rising
global income and making an approximation to the marginal abatement
costs (MAC) of CO\textsubscript{2}. Discounted mitigation expenditures
are shown to be a convex function of cumulative CO\textsubscript{2}
emissions. We also consider minimum-expenditure solutions for meeting
cumulative emissions goals, using a regularized variational method
yielding an initial value problem in the integrated decarbonization
rate. A quasi-stationary solution to this problem can be obtained
for a special case, yielding decarbonization rate that is proportional
to annual CO2 emissions. Minimum-expenditure trajectories in scenarios
where CO\textsubscript{2} emissions decrease must begin with rapid
decarbonization at rate decreasing with time. Due to the shape of
global MAC the fraction of global income spent on CO\textsubscript{2}
mitigation (\textquotedbl{}burden\textquotedbl{}) generally increases
with time, as cheaper avenues for mitigation are exhausted. Therefore
failure to rapidly decarbonize early on reduces expenditures by a
small fraction (on the order of 0.01 \%) of income in the present,
but leads to much higher burden to future generations (on the order
of 1 \% of income).
\end{abstract}

\section{Introduction}

Global warming from carbon-dioxide (CO\textsubscript{2}) is related
to its cumulative emissions, i.e. emissions integrated across time,
and independent of emissions pathway (\citet{Allen2009,Matthews2009,Stocker2013a,Seshadri2017}).
Mitigation of global warming requires large-scale and expensive efforts
to decarbonize the world economy (\citet{Manne1993,Grubb1993,Weyant1993,Nordhaus1993a}).
Estimated costs of mitigation vary across different studies and depend
on the assumptions that are made (\citet{Grubb1993,Rogelj2015}).
The Stern Review of Climate Change (\citet{Stern2007}) estimated
an average cost of reducing anthropogenic emissions of climate forcers
to a 550 ppm CO\textsubscript{2} equivalent stabilization level of
about 1 percent of Global Gross Domestic Product (GGDP), which is
much smaller than damage costs of unmitigated global warming (\citet{Stern2007}).
In a recent study \citet{Rogelj2015} illustrate using the integrated
assessment model MESSAGE that for scenarios limiting global warming
to less than 2 degrees C the cost of mitigation ranges from a small
fraction of 1\% to a few \% of GGDP, and an important factor governing
the costs is baseline energy demand (\citet{Rogelj2015}). 

This paper introduces an analytical model for global expenditures
on reducing CO\textsubscript{2} emissions from economic activity
(\textquotedbl{}decarbonization\textquotedbl{}). We represent the
marginal abatement cost (MAC) of reducing CO\textsubscript{2} emissions
as a function of emissions intensity of GGDP. If increase in GGDP
leads to relative increase in CO\textsubscript{2} emissions by a
smaller factor, because global income elasticity of CO\textsubscript{2}
emissions is less than $1$, there will occur exogenous reductions
in emissions intensity as the global economy grows; this will happen
even in the absence of any deliberate mitigation effort, in the \textquotedbl{}business
as usual\textquotedbl{} scenario. Exogenous reductions in emissions
intensity may be viewed as being the result of technological improvements
with time, in models where the production function is explicitly present
(\citet{Nordhaus1993}). Our model does not include the production
function, but instead uses the income elasticity of global CO\textsubscript{2}
emissions as a constant parameter. This permits a simple treatment
of the time-dependence of the MAC curve, even in the presence of exogenous
reductions in emissions intensity, but our discussion is limited by
the assumption of constant elasticity. 

We integrate the resulting MAC curve to consider global expenditures
on mitigation. Our model considers expenditures for reducing emissions
intensity (i.e. decarbonization) as well as those involved in scaling
up decarbonization activity as the global economy expands. This can
help understand factors behind the scale of economic effort involved
in global decarbonization. Because of the centrality of cumulative
CO\textsubscript{2} emissions to the global warming problem (\citet{Stocker2013a,Friedlingstein2014,Raupach2014,Rozenberg2015,Peters2016,Pfeiffer2016}),
we also consider how to minimize costs of limiting cumulative emissions
over time. This is examined through a constrained variational problem
(\citet{Brunt2004}), minimizing a functional describing discounted
total mitigation expenditures while satisfying a constraint on cumulative
emissions across a specified time-horizon. The policy variable is
the decarbonization rate, the latter being the rate with which emissions
intensity is reduced. Minimizing the functional leads to the familiar
Euler-Lagrange equation (\citet{Brunt2004}). For our model of CO\textsubscript{2}
mitigation expenditures, the variational problem is degenerate%
\footnote{See, for e.g., \citet{Brunt2004} for a general discussion of such
cases. %
} and the Euler-Lagrange equation is algebraic, whereas we seek an
initial condition problem in the integrated decarbonization rate.
This is an ill-posed problem (\citet{Tikhonov1977}), because a solution
satisfying the initial condition on the integrated decarbonization
rate does not exist for the algebraic Euler-Lagrange equation that
is obtained. One approach for dealing with ill-posed problems is through
regularization (\citet{Tikhonov1977}), by adding an additional term
to the quantity being minimized in order to render it soluble. In
our case we include an additional contribution to the functional being
minimized for rendering an initial value problem in the Euler-Lagrange
equation. 

The aforementioned approach contrasts with the optimal control problem
of choosing an emissions pathway to maximize discounted utility, taking
into account costs of mitigation as well as damage costs of global
warming, which underlies integrated assessment models of global warming
such as DICE (\citet{Nordhaus1993a,Nordhaus1993}). In such approaches
the optimal mitigation pathway is such that the effect on utility
of marginal increment to current consumption from increasing emissions
is balanced by the present value of the diminution in future consumption
(\citet{Nordhaus1993a}). Without considering the production function
or climate damages in the present paper we cannot model the above
features, and instead assume that cumulative emissions goals are specified
exogenously, following contemporary discussions in climate policy
(\citet{Meinshausen2009,Peters2016}). 

Section 2 introduces the models of global CO\textsubscript{2} emissions
and mitigation expenditures, and the results derived from them. Section
3 considers the expenditure-minimizing pathways of decarbonization,
subject to a cumulative emissions constraint. The control variable
is the integrated decarbonization rate. As mentioned previously, the
original problem is degenerate, and must be regularized to give rise
to an initial value problem in the integrated decarbonization rate.
Section 4 presents numerical illustrations of the main results of
the paper.

\section{Models}

\subsection{CO\protect\textsubscript{2} emissions under business as usual}

We describe global CO\textsubscript{2} emissions $m\left(t\right)$
as the product of GGDP, denoted by $g\left(t\right)$, and global
emissions intensity $\mu\left(t\right)$; with present values $m_{0}$,
$g_{0}$ and $\mu_{0}$. In the absence of deliberate mitigation,
under business as usual,%
\footnote{Absence of mitigation can also adversely impact economic growth. Quoting
\citet{Stern2016}, \textquotedbl{}So the business-as-usual baseline,
against which costs of action are measured, conveys a profoundly misleading
message to policymakers that there is an alternative option in which
fossil fuels are consumed in ever greater quantities without any negative
consequences to growth itself.\textquotedbl{} %
} growth in GGDP leads to increase in global CO\textsubscript{2} emissions
that is governed by the constant-elasticity model $\triangle m/m=\theta\triangle g/g$,
where $\theta$ is the global income elasticity of global CO\textsubscript{2}
emissions. Considering small time-interval $\triangle t$ the emissions
intensity at $t+\triangle t$ has formula%
\footnote{We write $\frac{m+\triangle m}{g+\triangle g}$ as $\mu\frac{1+\theta\frac{\triangle g}{g}}{1+\frac{\triangle g}{g}}$,
expand the denominator by its Taylor series in $\frac{\triangle g}{g}$
and approximate to first-degree in $\frac{\triangle g}{g}$, obtaining
equation (\ref{eq:1}).%
} 
\begin{equation}
\frac{m+\triangle m}{g+\triangle g}=\left(1+\left(\theta-1\right)\frac{\triangle g}{g}\right)\mu\label{eq:1}
\end{equation}
The change $\triangle\mu$ in emissions intensity is the above ratio
minus $\mu\left(t\right)$, so its rate of change during interval
$\triangle t$ is 
\begin{equation}
\frac{\triangle\mu}{\triangle t}=-\left(1-\theta\right)r\mu\label{eq:2}
\end{equation}
where $r=\left(\triangle g/\triangle t\right)/g$ is growth rate of
GGDP during this period. Generally $\theta$ is smaller than one,
leading to exogenous decrease of emissions intensity at rate $\sigma=\left(1-\theta\right)r$.
This effect is independent of any deliberate mitigation effort, occurring
under business as usual. It is larger for higher growth rates of GGDP
if $\theta<1$. For example, GGDP growth at constant rate $r=4.0$
\% (which is close to the historical mean during 1950-2014) and $\theta=0.75$
yields constant exogenous decarbonization rate of $1$ \% per year.
However we caution that income elasticity of energy demand varies
by country and is smaller for developed nations (\citet{Webster2008}).
Therefore it is liable to change with time as countries develop, and
our assumption of constant elasticity is only an idealization. 

Smaller GGDP growth rates would lead to smaller exogenous decarbonization
rates. If the rate of growth of GGDP were only $1.2$ \% then $\sigma=0.3$
\% per year. This paper considers only scenarios with constant growth
rate of GGDP, so that with constant $r$ and $\theta$ and in the
absence of deliberate reductions, the emissions intensity at time
$t$ would be $\mu\left(t\right)=\mu_{0}e^{-\sigma t}$, where $t=0$
denotes the present. Of course, if elasticity $\theta=1$ then there
is no exogenous decarbonization in our model, and decarbonization
occurs only through deliberate mitigation.

\subsection{Marginal abatement cost of mitigation}

Increase in GGDP leads to increases in emissions as described above
and corresponding expansion of mitigation possibilities, thereby stretching
horizontally the marginal abatement cost (MAC) curve as the global
economy grows. In the absence of deliberate reductions, this effect
is governed by exogenous decarbonization rate $\sigma$ so the MAC
curve expressed in terms of emissions intensity is 
\begin{equation}
C\left(\mu\left(t\right)\right)=\frac{\alpha}{\left(\frac{\mu\left(t\right)}{\mu_{0}e^{-\sigma t}}\right)^{\nu}}\label{eq:3}
\end{equation}
 where $C\left(\mu\left(t\right)\right)$ is cost of reducing emissions
intensity by one unit, for each unit of GGDP and $\nu>0$ is a constant
parameter. In case there is no deliberate mitigation so that $\mu\left(t\right)$
is equal to $\mu_{0}e^{-\sigma t}$ for all times then the MAC in
our model remains constant at $\alpha$, and thereby this model neglects
effects of learning.%
\footnote{A simple model for exogenous learning would make $\alpha$ a function
of time. %
} As deliberate mitigation proceeds thereby reducing ratio $\mu\left(t\right)/\mu_{0}e^{-\sigma t}$
the marginal cost increases. The MAC $C\left(\mu\left(t\right)\right)$
is in units of billion \$ / (Gton CO\textsubscript{2} year\textsuperscript{-1}).
The graph represented by $C\left(\mu\left(t\right)\right)$ has the
same scale as MAC curves described in \$ / ton CO\textsubscript{2}.

Parameter $\alpha$ describes the present MAC, in billion \$ / (Gton
CO\textsubscript{2} year\textsuperscript{-1}). The present cost
of reducing emissions intensity by $\triangle\mu$ in a year is $\alpha g_{0}\triangle\mu$.
The factor $g_{0}\triangle\mu$ has units of emissions (Gton CO\textsubscript{2}
year\textsuperscript{-1}). For decreasing emissions intensity between
time $t$ and $t+\triangle t$ by $\triangle\mu\left(t\right)$, the
mitigation expenditure is $C\left(\mu\left(t\right)\right)g\left(t\right)\triangle\mu\left(t\right)$.
GGDP is in units of trillion \$ / year, and global emissions intensity
in Gtons CO\textsubscript{2} / trillion \$, so expenditure is in
(billion \$ / (Gton CO\textsubscript{2} year\textsuperscript{-1})){*}(trillion
\$ / year){*}(Gton CO\textsubscript{2}/ trillion \$), or billions
of dollars.

\subsection{Effect of mitigation at rate $k\left(t\right)$ }

Deliberate reduction of emissions intensity, or mitigation, occurs
at rate $k\left(t\right)$, with the effect over time-interval $\triangle t$
being $\triangle\mu\left(t\right)=-k\left(t\right)\mu\left(t\right)\triangle t$
. In conjunction with exogenous reductions arising during economic
expansion in case $\theta<1$, the rate of change in emissions intensity
is $\triangle\mu\left(t\right)/\triangle t=-\left(k\left(t\right)+\sigma\right)\mu\left(t\right)$,
which is integrated for $\mu\left(t\right)=\mu_{0}e^{-\int_{0}^{t}k\left(s\right)ds}e^{-\sigma t}$.
We write this in terms of integrated decarbonization rate $K\left(t\right)=\int_{0}^{t}k\left(s\right)ds$,
so that $\mu\left(t\right)=\mu_{0}e^{-K\left(t\right)}e^{-\sigma t}$.
Decarbonization in the form of this integrated rate is the policy
variable in the optimization problem of Section 3.

\subsection{Mitigation expenditure}

The expenditure on mitigation has two contributions: initial cost
associated with reducing emissions intensity, and scaling up mitigation
to maintain reduced levels of emissions intensity as GGDP increases.

\subsubsection{Reducing emissions intensity}

Deliberate reductions in emissions intensity require expenditures
at levels described by the MAC curve. Reducing emissions intensity
by $\triangle\mu\left(t\right)$ between time $t$ and $t+\triangle t$
costs $C\left(\mu\left(t\right)\right)g\left(t\right)\left|\triangle\mu\left(t\right)\right|$.
Only the deliberate reduction in emissions intensity $\triangle\mu\left(t\right)=-k\left(t\right)\mu\left(t\right)\triangle t$
contributes to mitigation expenditures, so this contribution between
time $t$ and $t+\triangle t$ is $\alpha\left(\mu\left(t\right)/\left(\mu_{0}e^{-\sigma t}\right)\right)^{-\nu}g\left(t\right)k\left(t\right)\mu\left(t\right)\triangle t$,
with discounted sum across the specified time-period being 
\begin{equation}
E_{\mu}=\alpha\mu_{0}^{\nu}\triangle t\sum_{t}e^{-\delta t}e^{-\nu\sigma t}g\left(t\right)k\left(t\right)\left(\mu\left(t\right)\right)^{1-\nu}\label{eq:4}
\end{equation}
where $\delta$ is the rate of time-discounting. Substituting for
emissions intensity $\mu\left(t\right)=\mu_{0}e^{-K\left(t\right)}e^{-\sigma t}$
\begin{equation}
E_{\mu}=\alpha\mu_{0}\triangle t\sum_{t}e^{-\delta t}e^{-\sigma t}g\left(t\right)\dot{K}\left(t\right)e^{\left(\nu-1\right)K\left(t\right)}\label{eq:5}
\end{equation}

\subsubsection{Expansion of mitigation with growth in GGDP}

The second contribution to expenditures is due to scaling up of mitigation
for maintaining lower levels of emissions intensity as the global
economy expands. Between times $t$ and $t+\triangle t$ the activities
involved in reducing emissions intensity from $\mu_{0}e^{-\sigma t}$,
the value it would have in the absence of deliberate reductions, to
$\mu\left(t\right)$ that it actually has must be expanded proportionally
to increase in GGDP during this period. For each unit of GGDP, total
cost of reducing emissions intensity from $\mu_{0}e^{-\sigma t}$
to $\mu\left(t\right)$ at time $t$ is 
\begin{equation}
\int_{\mu_{0}e^{-\sigma t}}^{\mu\left(t\right)}\frac{\alpha}{\left(\mu/\left(\mu_{0}e^{-\sigma t}\right)\right)^{\nu}}\left(-d\mu\right)\label{eq:6}
\end{equation}
Time $t$ is fixed in the above equation so factor $e^{-\nu\sigma t}$
can be taken outside the integral. This contribution increases proportionally
with change in GGDP, so expenditure between time $t$ and $t+\triangle t$,
on scaling up mitigation when GGDP increases from $g\left(t\right)$
to $g\left(t\right)+\triangle g\left(t\right)$, is 
\begin{equation}
\frac{\alpha\mu_{0}^{\nu}e^{-\nu\sigma t}}{\nu-1}\left(\frac{1}{\mu\left(t\right)^{\nu-1}}-\frac{1}{\mu_{0}^{\nu-1}e^{-\left(\nu-1\right)\sigma t}}\right)\triangle g\left(t\right)\label{eq:7}
\end{equation}
and substituting for $\mu\left(t\right)$ as before the total discounted
expenditure becomes 
\begin{equation}
E_{g}=\frac{\alpha\mu_{0}}{\nu-1}\sum_{t}e^{-\delta t}e^{-\sigma t}\left(e^{\left(\nu-1\right)K\left(t\right)}-1\right)\triangle g\left(t\right)\label{eq:8}
\end{equation}

\subsubsection{Total expenditure}

The total discounted expenditure $E_{\mu}+E_{g}$ in continuous-time
is
\begin{equation}
E\left(t\right)=\int_{0}^{t}e^{-\delta s}P_{\mu}\left(s\right)ds+\int_{0}^{t}e^{-\delta s}P_{g}\left(s\right)ds\label{eq:9}
\end{equation}
where 
\begin{equation}
P_{\mu}\left(t\right)=\beta e^{-\sigma t}g\left(t\right)\dot{K}\left(t\right)e^{\left(\nu-1\right)K\left(t\right)}\label{eq:10}
\end{equation}
is annual expenditure from reducing emissions intensity, where $\beta=\alpha\mu_{0}$,
and 
\begin{equation}
P_{g}\left(t\right)=\frac{\beta}{\nu-1}e^{-\sigma t}\dot{g}\left(t\right)\left(e^{\left(\nu-1\right)K\left(t\right)}-1\right)\label{eq:11}
\end{equation}
is that from expansion of mitigation. Exogenous decarbonization has
the effect of decreasing both contributions to future expenditure
by $e^{-\sigma t}$. We therefore introduce the parameter $\rho=\sigma+\delta$,
combining its effect with that of discounting in time. 

In scenarios with constant growth rate of GGDP%
\footnote{Strictly, the model of constant exogenous decarbonization rate $\sigma$
assumes constant GGDP growth rate, so this is the implicit assumption
throughout the paper. %
} and constant mitigation rate $k$ the discounted expenditure then
becomes
\begin{equation}
E\left(t\right)=\beta g_{0}k\int_{0}^{t}e^{\left(\left(\nu-1\right)k+r-\rho\right)s}ds+\frac{\beta g_{0}r}{\nu-1}\int_{0}^{t}\left(e^{\left(\left(\nu-1\right)k+r-\rho\right)s}-e^{\left(r-\rho\right)s}\right)ds\label{eq:12}
\end{equation}
integrating to 
\begin{equation}
E\left(t\right)=\beta g_{0}k\frac{e^{\left(\left(\nu-1\right)k+r-\rho\right)t}-1}{\left(\nu-1\right)k+r-\rho}+\frac{\beta g_{0}r}{\nu-1}\left(\frac{e^{\left(\left(\nu-1\right)k+r-\rho\right)t}-1}{\left(\nu-1\right)k+r-\rho}-\frac{e^{\left(r-\rho\right)t}-1}{r-\rho}\right)\label{eq:13}
\end{equation}

In case of large mitigation rates and long time-horizons $\frac{e^{\left(\left(\nu-1\right)k+r-\rho\right)t}-1}{\left(\nu-1\right)k+r-\rho}\gg\frac{e^{\left(r-\rho\right)t}-1}{r-\rho}$,
so that 
\begin{equation}
E\left(t\right)\cong\beta g_{0}\left(k+\frac{r}{\nu-1}\right)\frac{e^{\left(\left(\nu-1\right)k+r-\rho\right)t}-1}{\left(\nu-1\right)k+r-\rho}\label{eq:14}
\end{equation}
 and the ratio of expenditures from expansion and reducing emissions
intensity is approximately $\frac{1}{\nu-1}\frac{r}{k}$. 

For short time-horizons,%
\footnote{We expand the corresponding exponentials in equation (\ref{eq:13})
by their Taylor series and consider the leading-order terms. %
} such that $\left(\left(\nu-1\right)k+r-\rho\right)t\ll1$, the expenditure
from reducing emissions intensity increases linearly with time as
$E_{\mu}\left(t\right)\cong\alpha m_{0}kt$, being proportional to
the decarbonization rate. The second contribution from expansion is
quadratic in time as $E_{g}\left(t\right)\cong\frac{\alpha m_{0}}{2}krt^{2}$,
and increases with the decarbonization rate and GGDP growth rate.
Their ratio is $E_{g}\left(t\right)/E_{\mu}\left(t\right)=\frac{1}{2}rt$,
and initially $rt\ll1$ so expenditures are governed by costs of reducing
emissions intensity, but the second contribution from expansion becomes
increasingly important.

\subsection{Burden of mitigation expenditure as fraction of GGDP}

The global economy of the future is expected to be richer than the
present, and therefore better poised to manage CO\textsubscript{2}
mitigation expenditures. However marginal abatement costs increase
with time as cheaper mitigation activities are exhausted and more
expensive activities must be undertaken for continued decarbonization.
Consider \textquotedbl{}burden\textquotedbl{}, defined as the ratio
of mitigation expenditure in a year and corresponding GGDP; the numerator
is the integrand in equation (\ref{eq:9}) without the time-discount
factor. The burden is 
\begin{equation}
b\left(t\right)=\beta e^{-\sigma t}k\left(t\right)e^{\left(\nu-1\right)K\left(t\right)}+\frac{\beta r}{\nu-1}e^{-\sigma t}\left(e^{\left(\nu-1\right)K\left(t\right)}-1\right)\label{eq:15}
\end{equation}
using $\dot{g}/g=r$. The two terms above arise from reducing emissions
intensity and expansion respectively. Let us consider their respective
contributions to $\dot{b}\left(t\right)$ for the case of constant
decarbonization rate $k$. 

From reducing emissions intensity this is
\begin{equation}
\dot{b}\left(t\right)=\beta k\left(\left(\nu-1\right)k-\sigma\right)e^{-\sigma t}e^{\left(\nu-1\right)kt}\label{eq:16}
\end{equation}
so its sign depends on the sign of $\left(\nu-1\right)k-\sigma$.
If the MAC rises steeply enough that $\nu>1$ then $\dot{b}\left(t\right)>0$
if the decarbonization rate is large enough compared to the exogenous
rate $\sigma$ so that $\left(\nu-1\right)k>\sigma$ . Increasing
burden to future generations can result from a sharply rising MAC
curve not being compensated adequately by exogenous reductions in
emissions intensity. In case $0<\nu<1$ then the burden from decarbonization
decreases with time. %
\footnote{In the approach $\nu\rightarrow1$ the contribution to decarbonization
is approximately $b\left(t\right)=\beta e^{-\sigma t}k\left(t\right)$,
so this generally decreases with time unless decarbonization rate
is increasing.%
} 

From expansion the contribution to $\dot{b}$ is simplified to 
\begin{equation}
\dot{b}\left(t\right)=\frac{\beta r}{\nu-1}\left\{ \left(\left(\nu-1\right)k-\sigma\right)e^{-\sigma t}e^{\left(\nu-1\right)kt}-\sigma e^{-\sigma t}\right\} \label{eq:17}
\end{equation}
and generally the last term is small so that if $\left(\left(\nu-1\right)k-\sigma\right)/\left(\nu-1\right)>0$
this contribution to burden increases with time. Only if $1<\nu<1+\sigma/k$
does this contribution decrease with time, and for all other conditions
it increases. If $\sigma/k\ll1$ the burden from expansion is almost
certain to increase with time because of the shape of the MAC curve. 

In summary, both decarbonization and expansion are likely to impose
increasing burdens on future generations if the MAC curve rises steeply
enough that 
\begin{equation}
\nu>1+\frac{\left(1-\theta\right)r}{k}\label{eq:18}
\end{equation}
since increasing MAC would not be compensated by effects of exogenous
reductions in emissions intensity. 

In the future if $e^{\left(\nu-1\right)K\left(t\right)}\gg1$ in equation
(\ref{eq:15}) the burden simplifies to 
\begin{equation}
b\left(t\right)=\beta e^{-\sigma t}e^{\left(\nu-1\right)K\left(t\right)}\left(k\left(t\right)+\frac{r}{\nu-1}\right)\label{eq:19}
\end{equation}
so for approximately constant values of integrated decarbonization
$K\left(t\right)$ the burden from reducing emissions intensity is
proportional to decarbonization rate $k\left(t\right)$.

\subsection{Convexity of relation between mitigation costs and cumulative emissions}

Here we examine the approximate relation between discounted mitigation
expenditures and cumulative CO\textsubscript{2} emissions. Such a
relation of course depends on the function $K\left(t\right)$, and
this subsection considers only scenarios with constant decarbonization
rate so $K\left(t\right)=kt$. We also fix the GGDP growth rate and
autonomous decarbonization rate, so the graph between expenditures
and cumulative emissions reflects only differences in the decarbonization
rate. 

Cumulative CO\textsubscript{2} emissions between the present at time
$t=0$ and the time-horizon at $t=T$ is $M\left(T\right)=\int_{0}^{T}m\left(t\right)dt$.
With $m\left(t\right)=\mu_{0}g\left(t\right)e^{-K\left(t\right)}e^{-\sigma t}$
, and in case of constant decarbonization rate $k$ this becomes $M\left(T\right)=m_{0}\left(1-e^{-\chi T}\right)/\chi$,
with $\chi=k+\sigma-r$. The slope of the graph between discounted
expenditure $E\left(T\right)$ and cumulative emissions $M\left(T\right)$
equals derivative $\partial E/\partial M$, which is written as $\partial E/\partial M=\left(\partial E/\partial k\right)\left(\partial k/\partial\chi\right)\left(\partial\chi/\partial M\right)$.
The value of $\partial M/\partial\chi=\frac{m_{0}}{\chi^{2}}\left(\left(1+\chi T\right)e^{-\chi T}-1\right)$.
We examine separately the very different cases with short and long
time-horizons. 

For short time-horizons the leading order terms are $\partial E/\partial k\cong\alpha m_{0}T$,
and $\partial M/\partial\chi\cong-m_{0}T^{2}$. Then, since $\partial k/\partial\chi=1$
with constant $r$ and $\sigma$, we obtain $\partial E/\partial M\cong-\alpha/T$.
This slope is constant for fixed time-horizon $T$. For $T=1$ the
slope of the graph is $-\alpha$ , which also represents the increase
in expenditure (in billion \$s) during one year from the present time
that is needed for decreasing emissions by $1$ Gton year\textsuperscript{-1}.
This much should be obvious from the model of the MAC curve, but there
is also an analogous linear relation for short time-horizons of a
few years.%
\footnote{The slope of the graph between expenditure and cumulative emissions
becomes smaller as time-horizon $T$ is increased because sensitivity
of expenditure to decarbonization rate grows linearly with $T$, whereas
that of cumulative emissions has a quadratic relation with $T$. The
benefits of higher decarbonization rates are sensitive to $T$, because
decarbonization takes time to influence cumulative emissions. Therefore
for time horizons of a few years, decarbonization appears more attractive
while considering the longer periods. %
} 

For long time-horizons $T$ for which $e^{-\chi T}\ll1$ , $M\cong m_{0}/\chi$,
and eliminating $k$ and substituting into equation (\ref{eq:14})
one obtains the relation between expenditure and cumulative emissions
$E\cong\beta g_{0}\frac{e^{\left(r-\rho\right)T}-1}{r-\rho}\left(\frac{m_{0}}{M}+\frac{r\nu}{\nu-1}-\sigma\right)$,
where $E$ is a convex function of $M$ and has increasing slope for
smaller $M$. For $\nu>1$ the expenditure in equation (\ref{eq:14})
depends more strongly on decarbonization rate $k$ due to the exponential
term in $k$, so the effect is even larger. Convexity of the graph
between $E$ and $M$ arises from two effects in the case of constant
decarbonization rates: $\partial M/\partial\chi$ for the case of
large $T$ behaves as $-m_{0}/\chi^{2}$, so larger $\chi$ in case
of more rapid decarbonization has progressively smaller effect on
limiting cumulative emissions; secondly, increasing decarbonization
rate increases $E$ nonlinearly for long time-horizons, due to the
exponential term in $k$. These effects are more general and therefore
not limited to scenarios with constant decarbonization rates. 

In summary, for short time-horizons the graph of mitigation expenditures
versus cumulative emissions is linear, whereas for longer time-horizons
it is convex. CO\textsubscript{2} is long-lived and policies to limit
cumulative emissions should take into account sufficiently long time-horizons.
Under these conditions the marginal cost, measured in terms of additional
discounted expenditures needed to achieve more stringent mitigation
goals, is increasing. 

Parameters used in the paper are listed in Table 1.

\pagebreak{}

Table 1: Description of parameters

\begin{tabular}{|c|c|c|}
\hline 
Symbol &
Description &
Units\tabularnewline
\hline 
\hline 
$m\left(t\right)$ &
CO\textsubscript{2} emissions &
Gton CO\textsubscript{2} year\textsuperscript{-1}\tabularnewline
\hline 
$M_{0}$ &
cumulative CO\textsubscript{2} emissions goal &
PgC\tabularnewline
\hline 
$\mu\left(t\right)$ &
CO\textsubscript{2} emissions intensity &
Gton CO\textsubscript{2} (trillion \$)\textsuperscript{-1}\tabularnewline
\hline 
$g\left(t\right)$ &
Global gross domestic product (GGDP) &
trillion \$ year\textsuperscript{-1}\tabularnewline
\hline 
$r$ &
annual GGDP growth rate &
year\textsuperscript{-1}\tabularnewline
\hline 
$k\left(t\right)$ &
decarbonization rate &
year\textsuperscript{-1}\tabularnewline
\hline 
$\theta$ &
income elasticity of CO\textsubscript{2} emissions &
dimensionless\tabularnewline
\hline 
$\sigma$ &
exogenous decarbonization rate &
year\textsuperscript{-1}\tabularnewline
\hline 
$\alpha$  &
coefficient for MAC curve &
billion \$ / (Gton CO\textsubscript{2} year\textsuperscript{-1})\tabularnewline
\hline 
$\nu$ &
exponent in MAC curve &
dimensionless\tabularnewline
\hline 
$E_{\mu}\left(t\right)$, $E_{g}\left(t\right)$, $E\left(t\right)$ &
discounted expenditures until year $t$ &
billion \$\tabularnewline
\hline 
$P_{\mu}\left(t\right)$, $P_{g}\left(t\right)$, $P\left(t\right)$ &
expenditure in year $t$ &
billion \$ year\textsuperscript{-1}\tabularnewline
\hline 
$b\left(t\right)$ &
expenditure / GGDP in year $t$ &
dimensionless\tabularnewline
\hline 
$K\left(t\right)$ &
integrated decarbonization rate &
dimensionless\tabularnewline
\hline 
$\delta$ &
time-discount rate &
year\textsuperscript{-1}\tabularnewline
\hline 
$\rho$ &
$\sigma+\delta$ &
year\textsuperscript{-1}\tabularnewline
\hline 
$\chi$ &
$k+\sigma-r$ &
year\textsuperscript{-1}\tabularnewline
\hline 
$\beta$ &
$\alpha\mu_{0}$ &
year\tabularnewline
\hline 
\end{tabular}

\pagebreak{}

\section{Minimum-expenditure pathways for reducing emissions intensity of
CO\protect\textsubscript{2}}

This section examines quasi-stationary pathways of decarbonization
that minimize mitigation expenditure subject to constraint on cumulative
CO\textsubscript{2} emissions. Such pathways need not correspond
to constant decarbonization rate. There is a constraint on the integrated
decarbonization rate, which is $K\left(0\right)=0$ at the present
time, and we therefore seek the initial value problem in $K\left(t\right)$
whose solution minimizes mitigation expenditure. 

Cumulative CO\textsubscript{2} emissions is written as $M\left(T\right)=\int_{0}^{T}m\left(t,K,\dot{K}\right)dt$,
where $m\left(t,K,\dot{K}\right)=\mu_{0}g\left(t\right)e^{-K\left(t\right)}e^{-\sigma t}$
is emissions.%
\footnote{Emissions do not depend explicitly on $\dot{K}$, but we include this
argument for consistency with the formulation of the rest of the functional
that we seek to minimize.%
} We wish to find $K\left(t\right)$ that minimizes discounted mitigation
expenditure $E\left(T\right)=\int_{0}^{T}f\left(t,K,\dot{K}\right)dt$,
where $f\left(t,K,\dot{K}\right)=\beta e^{-\delta t}e^{-\sigma t}g\left(t\right)\dot{K}\left(t\right)e^{\left(\nu-1\right)K\left(t\right)}+\frac{\beta}{\nu-1}e^{-\delta t}e^{-\sigma t}\dot{g}\left(t\right)\left(e^{\left(\nu-1\right)K\left(t\right)}-1\right)$. 

Consider choosing stationary pathway of integrated decarbonization
$K\left(t\right)$ in order to minimize $E\left(T\right)$ subject
to cumulative emissions constraint 
\begin{equation}
\int_{0}^{T}m\left(t,K,\dot{K}\right)dt=M_{0}\label{eq:20}
\end{equation}
where $M_{0}$ is the cumulative emissions goal. For such a pathway,
the derivative of functional $I\left(K,\dot{K}\right)=\int_{0}^{T}f\left(t,K,\dot{K}\right)dt+\lambda_{1}\left\{ \int_{0}^{T}m\left(t,K,\dot{K}\right)dt-M_{0}\right\} $
must be stationary with respect to $K\left(t\right)$ for arbitrary
perturbations $\delta K$ satisfying the initial condition. This yields
the familiar Euler-Lagrange (E-L) equation derived in Appendix 1 
\begin{equation}
\frac{\partial f}{\partial K}+\lambda_{1}\frac{\partial m}{\partial K}=\frac{d}{dt}\left(\frac{\partial f}{\partial\dot{K}}+\lambda_{1}\frac{\partial m}{\partial\dot{K}}\right)\label{eq:21}
\end{equation}
simplifying to $e^{\nu K\left(t\right)}=\frac{\lambda_{1}\mu_{0}}{\left(\delta+\sigma\right)\beta}e^{\delta t}$,
where $\lambda_{1}$ can be eliminated using the constraint on cumulative
emissions. In addition the solution must satisfy a \textquotedbl{}natural
boundary condition\textquotedbl{} $\frac{\partial f}{\partial\dot{K}}\left(T\right)+\frac{\partial m}{\partial\dot{K}}\left(T\right)=0$
arising from the fact that the value of $K\left(T\right)$ is not
fixed by the specification of our problem (Appendix 1). However such
satisfaction is not possible, and we seek a solution that is only
stationary with respect to perturbations that leave intact not only
$K\left(0\right)$, which follows from initial condition $K\left(0\right)=0$,
but also $K\left(T\right)$. Fixing $K\left(T\right)$ is an artificial
constraint on our problem that has been introduced for tractability.
In this sense our solution is \textquotedbl{}quasi-stationary\textquotedbl{},
i.e. only relative to a restricted type of perturbations which preserves
both endpoints, although the second endpoint is not constrained by
a final condition on $K\left(t\right)$. 

Although $K\left(t\right)$ increases in time in the presence of a
non-zero discount rate, the absence of a term in $\dot{K}$ in the
E-L equation precludes imposing initial condition $K\left(0\right)=0$.
The E-L equation is algebraic in our optimization problem because
integrand $f+\lambda_{1}m$ depends linearly on $\dot{K}$, leading
to a degenerate case (\citet{Brunt2004}) as shown in Appendix 1. 

In order to introduce a term in $\dot{K}$ in the E-L equation, we
seek a second integral constraint involving a different function $h\left(t,K,\dot{K}\right)dt$
that obeys
\begin{equation}
\frac{d}{dt}\left(\frac{\partial h}{\partial\dot{K}}\right)=\dot{K}e^{-\gamma t}\label{eq:22}
\end{equation}
with $\gamma>0$, for reasons that will become evident. Then $\frac{\partial h}{\partial\dot{K}}=\int_{0}^{t}\dot{K}\left(s\right)e^{-\gamma s}ds$
and integrating by parts 
\begin{equation}
\frac{\partial h}{\partial\dot{K}}=e^{-\gamma t}K\left(t\right)+\gamma\int_{0}^{t}K\left(s\right)e^{-\gamma s}ds\label{eq:23}
\end{equation}
using initial condition $K\left(0\right)=0$. Furthermore, choosing
$\gamma$ large so that the first term can be neglected we obtain
\begin{equation}
h\left(t,K,\dot{K}\right)=\gamma\dot{K}\left(t\right)\int_{0}^{t}K\left(s\right)e^{-\gamma s}ds+h_{1}\left(K\left(t\right),t\right)\label{eq:24}
\end{equation}
We seek only one such function parameterized by $\gamma$ satisfying
equation (\ref{eq:22}), so make the simplest choice and set $h_{1}\left(K\left(t\right),t\right)$
to zero. Then $h\left(t,K,\dot{K}\right)=\gamma\dot{K}\left(t\right)\int_{0}^{t}K\left(s\right)e^{-\gamma s}ds$,
and we choose $\gamma$ large so that $\int_{0}^{T}h\left(t,K,\dot{K}\right)dt$
can be made small.%
\footnote{To see how this is possible, consider the example of constant decarbonization
rate where $K\left(t\right)=\zeta t$ with $\zeta>0$, so that $h\left(t\right)=\gamma\zeta^{2}\int_{0}^{t}se^{-\gamma s}ds$.
Integrating by parts this becomes $-\zeta^{2}te^{-\gamma t}+\frac{\zeta^{2}}{\gamma}\left(1-e^{-\gamma t}\right)$,
which can be made small by choosing $\gamma$ to be sufficiently large.
The terms involved in defining $h\left(t\right)$ are positive so
$h\left(t\right)>0$ and furthermore in this example $h\left(t\right)<\zeta^{2}/\gamma$,
so $\int_{0}^{T}h\left(t\right)dt<\zeta^{2}T/\gamma$, which can be
made small by choosing $\gamma$ sufficiently large. %
} 

We therefore impose a further equality constraint on our original
problem
\begin{equation}
\int_{0}^{T}h\left(t,K,\dot{K}\right)dt=\varepsilon\label{eq:25}
\end{equation}
with $\varepsilon\ll1$ and the new functional to be minimized contains
an additional term $\lambda_{2}\left\{ \int_{0}^{T}h\left(t,K,\dot{K}\right)dt-\varepsilon\right\} $,
so the modified E-L equation becomes 
\begin{equation}
\frac{\partial f}{\partial K}+\lambda_{1}\frac{\partial m}{\partial K}+\lambda_{2}\frac{\partial h}{\partial K}=\frac{d}{dt}\left(\frac{\partial f}{\partial\dot{K}}+\lambda_{1}\frac{\partial m}{\partial\dot{K}}+\lambda_{2}\frac{\partial h}{\partial\dot{K}}\right)\label{eq:26}
\end{equation}
Here too there is an analogous natural boundary condition, but we
cannot satisfy it and only seek a solution that is stationary with
respect to the restricted class of perturbations discussed above. 

Recall that $\frac{d}{dt}\left(\frac{\partial h}{\partial\dot{K}}\right)=\dot{K}e^{-\gamma t}$.
Furthermore, $\frac{\partial h}{\partial K}=\gamma\dot{K}\left(t\right)\int_{0}^{t}e^{-\gamma t}dt=\dot{K}\left(1-e^{-\gamma t}\right)$.
Then the E-L equation is
\begin{equation}
-\lambda_{1}\mu_{0}g\left(t\right)e^{-\sigma t}e^{-K\left(t\right)}+\lambda_{2}\dot{K}\left(t\right)\left(1-e^{-\gamma t}\right)=-\left(\delta+\sigma\right)\beta e^{-\left(\delta+\sigma\right)t}g\left(t\right)e^{\left(\nu-1\right)K\left(t\right)}+\lambda_{2}\dot{K}\left(t\right)e^{-\gamma t}\label{eq:27}
\end{equation}
We simplify by neglecting $e^{-\gamma t}$ compared to $1$ because
$\gamma t\gg1$, so the evolution equation for $K$ becomes 
\begin{equation}
\dot{K}\left(t\right)=\frac{\lambda_{1}\mu_{0}}{\lambda_{2}}e^{-\sigma t}g\left(t\right)e^{-K\left(t\right)}-\frac{\left(\delta+\sigma\right)\beta}{\lambda_{2}}e^{-\left(\delta+\sigma\right)t}g\left(t\right)e^{\left(\nu-1\right)K\left(t\right)}\label{eq:28}
\end{equation}
with initial condition $K\left(0\right)=0$. Multiplying by $e^{K\left(t\right)}$
on both sides and defining $x\left(t\right)=e^{K\left(t\right)}$
the evolution equation for $x\left(t\right)$ is
\begin{equation}
\dot{x}\left(t\right)=\frac{\lambda_{1}\mu_{0}}{\lambda_{2}}e^{-\sigma t}g\left(t\right)-\frac{\left(\delta+\sigma\right)\beta}{\lambda_{2}}e^{-\left(\delta+\sigma\right)t}g\left(t\right)\left(x\left(t\right)\right)^{\nu}\label{eq:29}
\end{equation}
with $x\left(0\right)=1$. The above equation is not of a standard
type that can be solved exactly, and we resort to approximate solutions
(Appendix 2). Integrated decarbonization rate $K\left(t\right)$ must
increase to be economically relevant, but with an initial value problem
in $K\left(t\right)$ we face the risk that it actually decreases
in the solution to the above equation. In case of either $\sigma>0$
or $\delta>0$ this possibility is realized and, as shown in Appendix
2 for $\sigma>0$, $K\left(t\right)$ is a decreasing function. We
cannot avoid this by regularizing the E-L equation for a 2-point boundary
value problem. The value of $K\left(0\right)$ is known but not $K\left(T\right)$,
and the latter is not uniquely determined by the constraint on cumulative
emissions.%
\footnote{The cumulative emissions constraint of equation (\ref{eq:20}) discretized
in time corresponds to a single equation in several unknown values
of $K$ at the various time-steps, so this does not yield a unique
constraint on $K\left(T\right)$. %
} 

Therefore we can obtain a meaningful solution to the E-L equation
only for the special case of $\sigma=0$, involving unit income elasticity
of emissions, and in the absence of time-discounting. For this special
case, $x\left(t\right)=1+\frac{\lambda_{1}\mu_{0}}{\lambda_{2}}G\left(t\right)$
, where $G\left(t\right)=\int_{0}^{t}g\left(s\right)ds$ is integrated
GGDP, and hence the quasi-stationary solution is 
\begin{equation}
K\left(t\right)=\ln\left(1+\frac{\lambda_{1}\mu_{0}}{\lambda_{2}}G\left(t\right)\right)\label{eq:30}
\end{equation}
which is increasing. Lagrange multipliers $\lambda_{1}$ and $\lambda_{2}$
are estimated by substituting equation (\ref{eq:30}) into integral
constraints provided by equations (\ref{eq:20}) and (\ref{eq:25}).
The E-L equation has reduced our infinite-dimensional problem of choosing
function $K\left(t\right)$ to the finite-dimensional one of estimating
$\lambda_{1}$ and $\lambda_{2}$ satisfying these constraints. 

For the above solution, being quasi-stationary only if $\sigma=0$,
we have $e^{-K\left(t\right)}=1/\left(1+\frac{\lambda_{1}\mu_{0}}{\lambda_{2}}G\left(t\right)\right)$;
and therefore in this case cumulative emissions at time $t$ is $M\left(t\right)=\frac{\lambda_{2}}{\lambda_{1}}\ln\left(1+\frac{\lambda_{1}\mu_{0}}{\lambda_{2}}G\left(t\right)\right)$,
or $M\left(t\right)=\frac{\lambda_{2}}{\lambda_{1}}K\left(t\right)$
in the quasi-stationary solution. 

Thus the quasi-stationary solution, stationary relative to a restricted
class of perturbations in $K$, exists for the special case of $\sigma=0$
and $\delta=0$, and has cumulative emissions graph proportional to
the graph of $K\left(t\right)$, and the emissions graph $m\left(t\right)$
proportional that of decarbonization rate $k\left(t\right)$, with
$k\left(t\right)=\frac{\lambda_{1}}{\lambda_{2}}m\left(t\right)$.
Scenarios requiring decreasing emissions, so that cumulative emissions
eventually becomes approximately constant, have $K\left(t\right)$
increasing at diminishing rate in this solution; decarbonization rate
$k\left(t\right)$ is initially large and decreases with time.

\section{Numerical Results}

\subsection{Parameter estimates and scenarios}

We estimate the global MAC curve (in 1990 US dollars) for CO\textsubscript{2}
by aggregating estimates presented in \citet{Morris2008}, which in
turn are based on the MIT Emissions Prediction and Policy Analysis
(EPPA) model (\citet{Paltsev2005}). Figure 1a shows results from
\citet{Morris2008} for the year 2050. Estimation of the model of
MAC is illustrated in Figure 1b, and we must know the reference emissions
in the business as usual case, corresponding to the effect of $\mu_{0}e^{-\sigma t}$.
This is taken from reference-case emissions in 2050 documented for
the EPPA model in \citet{Paltsev2005}. Least squares regression yields
estimates for $\alpha$ and $\mu$ (Table 2). The exponent $\nu$
is significantly larger than one, and this relation affects the properties
examined in Section 2. There is large uncertainty in MACs (\citet{Criqui1999,Klepper2004,Amann2009})
but it appears that this exponent in our model cannot be much smaller
than $2$, because that would lead to very slow increase of the MAC
as decarbonization proceeds. Figure 1c shows the effect of changing
$\nu$ in our model. In case $\nu$ is closer to $1$ then the MAC
even after $50$ \% reductions compared to the reference case would
be less than $50$ \$ /tonne in 2005 prices, which is substantially
smaller than studies suggest (\citet{Ellerman1998,Klepper2004,Amann2009}). 

For income elasticity of CO\textsubscript{2} emissions, we use constant
value of $\theta=0.75$ yielding exogenous decarbonization rate of
$\sigma=1$ \% per year in a $4$ \% GGDP growth scenario that is
close to the recent historical value in real dollars (\citet{DeLong1998}).
As a result the estimate of $\sigma$ corresponds to the 2015 value
in the DICE model (\citet{Nordhaus2013}). In DICE the autonomous
decarbonization rate is decreasing at $0.1$\% at each time-step of
five years (\citet{Nordhaus2013}). If were were to consider scenarios
with decreasing GGDP growth rate, a similar decrease in exogenous
decarbonization rate would occur in our emissions model. 

For future GGDP growth estimates, considering that our time-horizon
is $100$ years from the present, most IAMs show a gradual decrease
in GGDP growth reflecting demographic transitions during the present
century (e.g. \citet{Paltsev2005}). Nevertheless we consider annual
growth rates (in real dollars) ranging between $0.012$ - $0.036$,
which is a larger range than in IPCC\textquoteright s Special Report
on Emissions Scenarios (SRES) (\citet{Krakauer2014}). Generally there
is a tendency even among experts to underestimate uncertainty (\citet{Morgan1990}),
and we included a wide range of long-term growth rates in our calculations.%
\footnote{The same caution applies to long term MACs but we do not consider
that effect here. %
} Estimates of parameters of the model, including present emissions,
GGDP, and emissions intensity, are listed in Table 2. 

We study cumulative emissions goals during the next $100$ years of
$300$, $600$, $900$ and $1200$ PgC.%
\footnote{Only the cumulative emissions are specified in PgC, because carbon
accounting is usually performed in these units. However the economic
model is carried out in units of Gton CO\textsubscript{2}. %
} Global warming is approximately proportional to cumulative CO\textsubscript{2}
emissions, with the ratio estimated to vary between $0.8-2.5$ K per
$1000$ PgC (\citet{Allen2014}). Assuming a symmetric distribution
of this \textquotedbl{}transient climate response to cumulative carbon
emissions\textquotedbl{} with mean value of $1.65$ K per $1000$
PgC yields mean global warming contribution of CO\textsubscript{2}
of approximately $0.5$, $1.0$, $1.5$, and $2.0$ K respectively
during the next $100$ years. With present global warming of about
$1$ K, if other contributions remain unchanged, the above cumulative
emissions goals correspond roughly to mean forecasts of global warming
of $1.5$, $2.0$, $2.5$, and $3.0$ K relative to preindustrial
conditions. We should emphasize that there are uncertainties in the
relation between cumulative emissions and global warming (\citet{Meinshausen2009,Peters2016}). 

\pagebreak{}

Table 2: Parameter estimates for model of expenditures. Monetary units
refer to 1990 USD. 

\begin{tabular}{|c|c|c|}
\hline 
Parameter &
Value &
Unit\tabularnewline
\hline 
\hline 
$\alpha$ &
$10.4\pm2.7$ &
billion \$ / (Gton CO\textsubscript{2} year\textsuperscript{-1})\tabularnewline
\hline 
$\nu$ &
$2.4\pm0.46$ &
dimensionless\tabularnewline
\hline 
$\theta$ &
$0.75$ &
dimensionless\tabularnewline
\hline 
$m_{0}$ &
$36$ &
Gton CO\textsubscript{2} year\textsuperscript{-1}\tabularnewline
\hline 
$\mu_{0}$ &
$0.46$ &
Gton CO\textsubscript{2} (trillion \$)\textsuperscript{-1}\tabularnewline
\hline 
$g_{0}$ &
$77.8$ &
trillion \$ year\textsuperscript{-1}\tabularnewline
\hline 
$\beta$ &
$4.8\times10^{-3}$ &
year\tabularnewline
\hline 
\end{tabular}

\pagebreak{}

\begin{figure}
\includegraphics[scale=0.8]{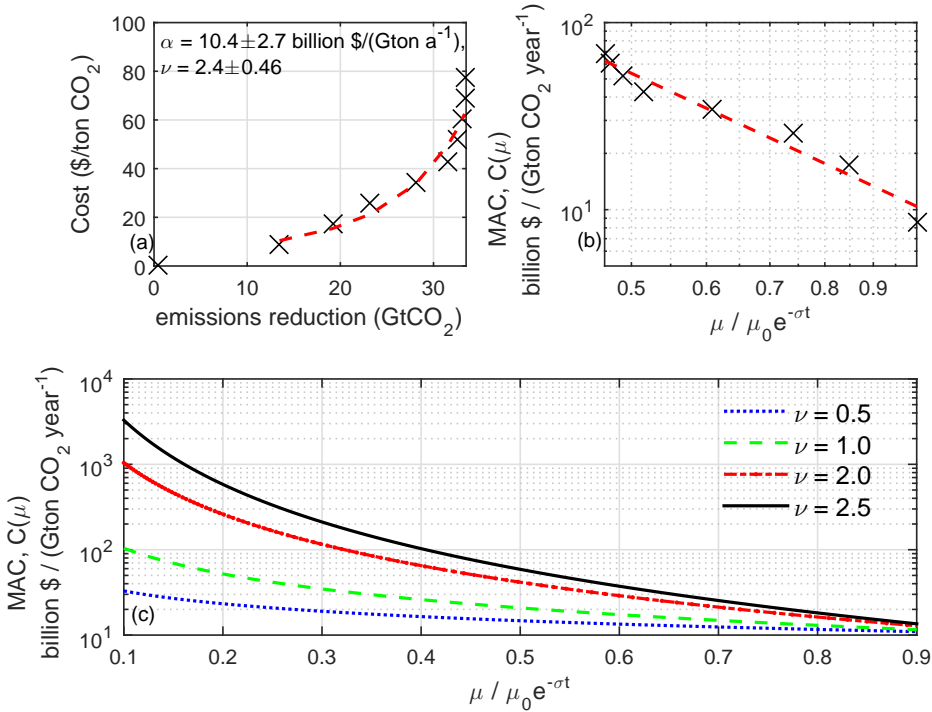}

\protect\caption{Estimate of MAC (in 1990 USD) and effect of changing exponent $\nu$
in our model of MAC: (a) crosses show global MAC versus quantity of
emissions reductions from \citet{Morris2008} for the year 2050, after
conversion from 2005 to 1990 dollars; (b) results of panel (a) presented
in the form of the model in equation (\ref{eq:3}), where $\mu_{0}e^{-\sigma t}$
is taken to correspond to a 60\% increase in 2050 CO\protect\textsubscript{2}
emissions compared to the present in the reference scenario, roughly
based on results documented for the EPPA integrated assessment model
(\citet{Paltsev2005}). The power law is estimated from a least squares
fit using the logarithm of the ordinate, yielding $\alpha=10.4$ billion
\$ / (Gton CO\protect\textsubscript{2} year\protect\textsuperscript{-1})
and $\nu=2.4$. In both panels, curves correspond to estimate of equation
(\ref{eq:3}); (c) effect of changing $\nu$ in the model, keeping
$\alpha$ constant. Values of $\nu$ much smaller than $2$ lead to
very slow increase of MAC (see text). }
\end{figure}

\subsection{Expenditure-minimizing decarbonization pathways in case of unit income
elasticity of emissions}

Figure 2 shows expenditure-minimizing pathways in case global income
elasticity of emissions is unity, so there is no exogenous decarbonization.
Recall that these are \textquotedbl{}quasi-stationary\textquotedbl{},
i.e. stationary relative to a restricted type of perturbations in
the variable $K\left(t\right)$ that keep the endpoints as fixed.
The quasi-stationary pathway of annual CO\textsubscript{2} emissions
is approximately invariant of GGDP growth rate. The time-dependence
of emissions arises from term $e^{rt-\int_{0}^{t}k\left(s\right)ds}$,
and larger $r$ is being compensated by larger $k$. More rapid economic
growth requires faster decarbonization for limiting cumulative emissions
to the same levels. 

As noted in Section 3, the quasi-stationary solution has $k\left(t\right)\propto m\left(t\right)$
so the decarbonization rate is constant only in scenarios involving
constant emissions, and increasing in time in scenarios involving
increasing emissions. For scenarios requiring decreasing emissions,
the decarbonization rate is initially larger and decreases with time.
Large annual decarbonization rates of more than $10$ percent are
initially involved in the quasi-stationary solution for limiting cumulative
emissions to $300$ PgC or less. 

\begin{figure}
\includegraphics[scale=0.8]{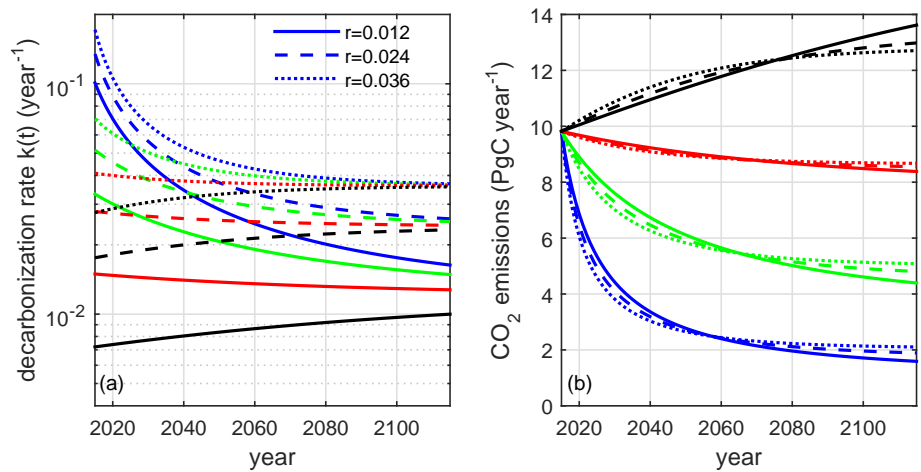}

\protect\caption{Quasi-stationary decarbonization rates $k\left(t\right)$ in case
$\theta=1$ and $\delta=0$, for cumulative emissions goals during
the next $100$ years of $300$, $600$, $900$ and $1200$ PgC in
the different colors, and different GGDP growth rates $r$: (a) decarbonization
rate $k\left(t\right)$, on a logarithmic scale; (b) annual CO\protect\textsubscript{2}
emissions. Where exogenous decarbonization is absent, i.e. $\sigma=0$,
the quasi-stationary emissions trajectory is approximately invariant
of GGDP growth rate. The decarbonization rate is proportional to emissions
as $k\left(t\right)=\frac{\lambda_{1}}{\lambda_{2}}m\left(t\right)$
and scenarios involving emissions reduction have decarbonization rate
decreasing with time. Solid, dashed, and dotted curves indicate GGDP
growth rate $r$ of $0.012$, $0.024$, and $0.036$ respectively. }
\end{figure}

\subsection{Mitigation expenditures in the presence of exogenous decarbonization}

Henceforth we will only describe cases with income elasticity of emissions
$\theta=0.75$, so exogenous decarbonization occurs at a rate increasing
with GGDP growth rate. Figure 3 shows pathways satisfying equation
(\ref{eq:30}) with the cumulative emissions constraint taking into
account exogenous decarbonization at rate $\sigma$. Limiting cumulative
emissions below $600$ PgC requires near-term decarbonization rates
of at least a few percent. Emissions pathways are generally not invariant
of GGDP growth rate in the presence of autonomous decarbonization,
and larger early emissions in high growth scenarios is compensated
by reduced emissions later on. When autonomous decarbonization is
present, it is not necessary that decreasing emissions pathways be
accompanied by initially high decarbonization rates that subsequently
decrease. However even in this case stringent mitigation scenarios
generally involve initially larger decarbonization rates. 

Despite decreasing decarbonization rates in such scenarios, mitigation
expenditures are rising here as in all mitigation scenarios, as Figure
4 shows. Expenditures increase from a few to tens of billion dollars
in the present time to a few orders of magnitude more as the MAC rises
and GGDP increases. The carbon price needed to achieve these reductions
in CO\textsubscript{2} is equal to the MAC, with formula $\alpha e^{\nu K\left(t\right)}$
following equation (\ref{eq:3}). Figure 4a shows the carbon price
(in constant 1990 US dollars) increasing to 100 \$ / ton CO\textsubscript{2}
in a few decades in the 300 PgC scenario, and much more rapidly in
case GGDP growth can expected to be rapid, because decarbonization
must occur more quickly. However the average rate of growth of GGDP
during the next 100 years will not be known until the end of this
period. 

In our model of mitigation cost, there are two contributions: from
reducing emissions intensity, and from subsequent expansion of mitigation
as the global economy grows. Recall from Section 2.4 that the first
contribution initially grows linearly in time, whereas the second
grows quadratically. The first contribution is initially much larger,
but the second contribution grows in importance with time (Figure
4d). Its role increases with time because, as the economy grows, decarbonization
efforts have to be scaled up in proportion to the level of decarbonization
at that time present the global economy, as measured by $e^{\left(\nu-1\right)K\left(t\right)}-1$.
The ratio of the two contributions is $P_{g}\left(t\right)/P_{\mu}\left(t\right)=\frac{1}{\nu-1}\frac{r}{k\left(t\right)}\left(1-e^{-\left(\nu-1\right)K\left(t\right)}\right)$.
For large $t$ this approaches $\frac{1}{\nu-1}\frac{r}{k\left(t\right)}$.
Since $\nu\cong2$ in our model, this always remains smaller than
one in scenarios where the decarbonization rate must be larger than
the GGDP growth rate. 

One lesson from Figures 2-4 is the importance of assumptions about
future global economic growth to estimates of the scale of effort
involved in decarbonization. Faster growth requires further decarbonization,
entailing not only higher marginal costs, but also these must be scaled
up to a larger extent with greater economic expansion. Uncertainty
in future economic growth thereby presents substantial uncertainty
for estimates of the scale of mitigation required. 

\begin{figure}
\includegraphics[scale=0.8]{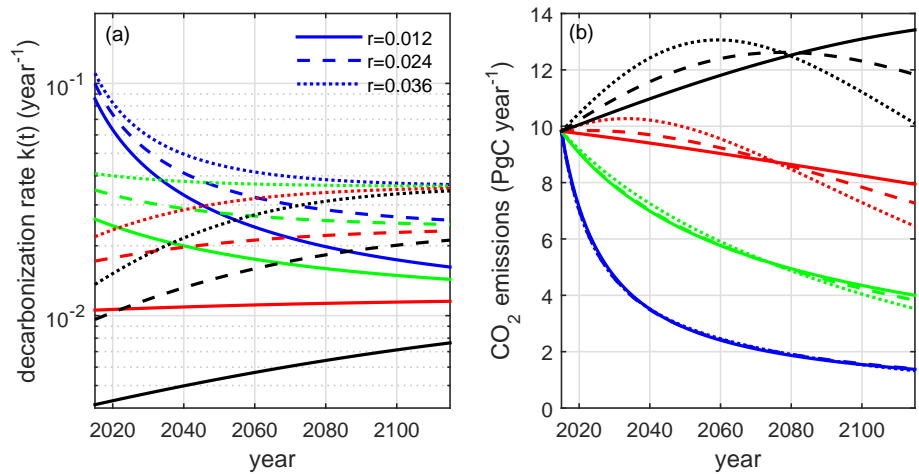}

\protect\caption{Quasi-stationary decarbonization rates for limiting cumulative emissions
during the next $100$ years to $300$, $600$, $900$ and $1200$
PgC in the different colors, where effects on emissions of exogenous
decarbonization arising from income elasticity of emissions $\theta=0.75$
have been included: (a) decarbonization rate $k\left(t\right)$; (b)
annual CO\protect\textsubscript{2} emissions. Solid, dashed, and
dotted curves indicate GGDP growth rate $r$ of $0.012$, $0.024$,
and $0.036$ respectively. }
\end{figure}

\begin{figure}
\includegraphics[scale=0.85]{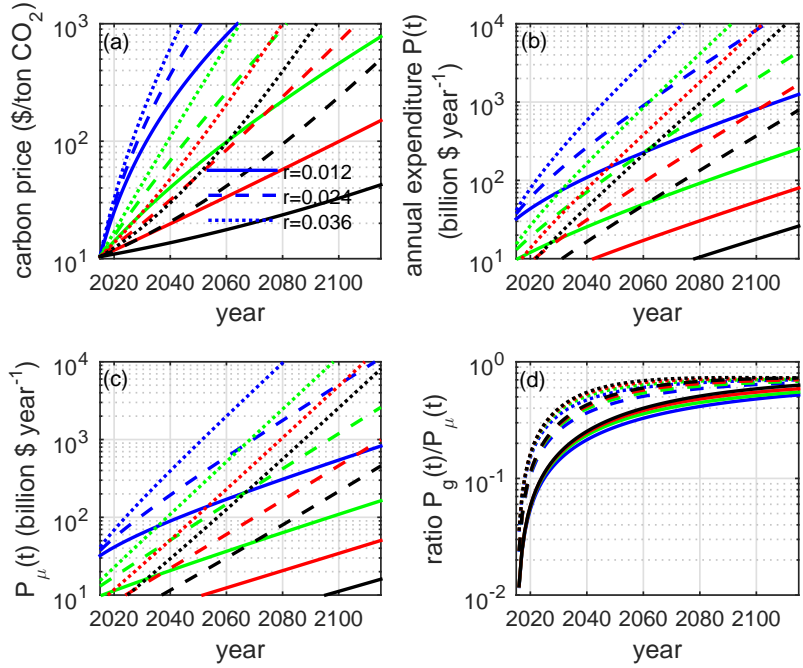}

\protect\caption{Carbon price and CO\protect\textsubscript{2} mitigation expenditures
for the scenarios in Figure 3. Colors indicate cumulative emissions
goals of $300$, $600$, $900$ and $1200$ PgC during the next 100
years. Solid, dashed, and dotted curves indicate GGDP growth rate
$r$ of $0.012$, $0.024$, and $0.036$ respectively. Prices and
expenditures are in constant 1990 USD: (a) carbon price, which is
equal to the marginal abatement cost; (b) total expenditure in each
year $P\left(t\right)$; (c) expenditure from reducing emissions intensity
$P_{\mu}\left(t\right)$; (d) ratio $P_{g}\left(t\right)/P_{\mu}\left(t\right)$
of expenditures from expansion and reducing emissions intensity. Keeping
cumulative emissions below 600 PgC typically requires carbon price
of 50 USD (in 1990 dollars) or higher within the next two decades.
Carbon price and expenditures are much higher in case of higher GGDP
growth. Mitigation expenditures are increasing even as decarbonization
rate is decreasing, as the MAC rises. At the present time expenditures
are dominated by those on reducing emissions intensity, but the contribution
from expansion plays an increasing role. }
\end{figure}

\subsection{Mitigation burden}

Figure 5 plots the mitigation expenditure as fraction of GGDP (\textquotedbl{}burden\textquotedbl{})
for the cases shown in the previous figures. This generally increases
with time because the condition of equation (\ref{eq:18}) is met,
and the effect of increasing MAC is not mitigated by exogenous decarbonization.
For short times the second contribution from expansion is small and
the burden approximates to $b\left(t\right)\cong\beta e^{\left(\nu-1\right)K\left(t\right)-\sigma t}k\left(t\right)$,
and with $\beta=4.8\times10^{-3}$ years in the present model this
becomes a very small fraction of GGDP even for large decarbonization
rates. At the current time there are many low-cost options for decarbonization
and the mitigation burden therefore can be very small. 

The burden after a long time is $b\left(t\right)\cong\beta e^{\left(\nu-1\right)K\left(t\right)-\sigma t}\left(k\left(t\right)+\frac{r}{\nu-1}\right)$.
The exponent $\left(\nu-1\right)K\left(t\right)-\sigma t$ increases
substantially as mitigation proceeds (Figure 5a) and the burden approaches
1 \% of GGDP or more in scenarios with stringent mitigation goals,
especially in the presence of rapid GGDP growth (Figure 5b). The latter
is generally consistent with results of \citet{Rogelj2015} showing
the large influence of baseline energy demand on discounted mitigation
costs. 

One effect is that delaying decarbonization can impose much larger
burdens on future generations if they seek to meet the same stringent
cumulative emissions goal. Figure 6 considers emissions scenarios
leading to cumulative emissions of $300$ PgC. We compare, in the
presence of exogenous decarbonization, the aforementioned quasi-stationary
scenario with one having constant decarbonization rate. The former
serves as archetype for rapid and early mitigation. A small fraction
($\leq$ $0.1$ \%) of GGDP is saved in the present time by choosing
the constant mitigation rate scenario, but this would entail much
higher mitigation burdens in the future. Towards the end of our time-horizon,
higher expenditures amount to about $1$\% of GGDP, and much more
in scenarios involving moderate to high growth. Present savings are
much more than offset by future increases in mitigation expenditure,
because future emissions must be correspondingly smaller in case of
failure to reduce emissions substantially in the present. A delay
in mitigation requires higher rates of decarbonization in the future
under conditions of much higher MAC, where the factor $e^{\left(\nu-1\right)K\left(t\right)-\sigma t}$
is much larger. The larger future decarbonization rates occurring
in the scenario with late mitigation are therefore amplified by a
much greater amount than present savings. 

\begin{figure}
\includegraphics[scale=0.8]{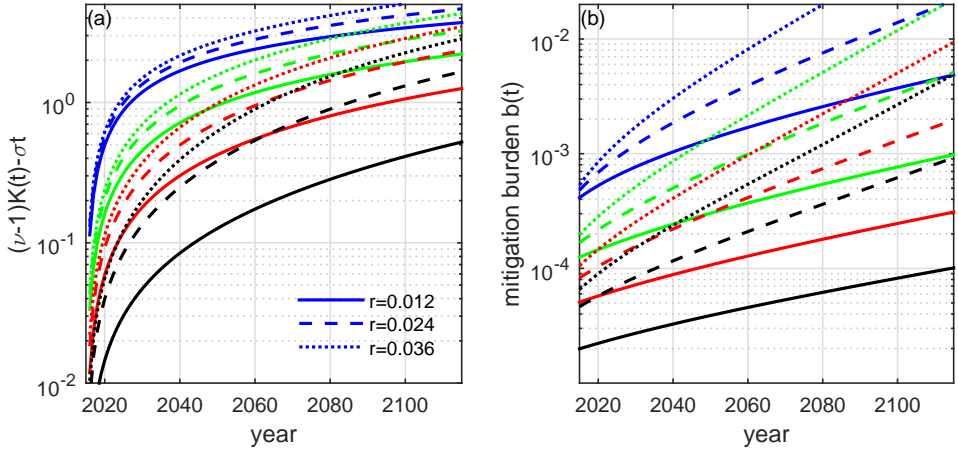}

\protect\caption{Mitigation expenditure as a fraction of GGDP (\textquotedbl{}burden\textquotedbl{})
in the scenarios of Figure 3. Colors indicate cumulative emissions
goals of $300$, $600$, $900$ and $1200$ PgC for the next 100 years.
Solid, dashed, and dotted curves indicate GGDP growth rate $r$ of
$0.012$, $0.024$, and $0.036$ respectively: (a) exponent$\left(\nu-1\right)K\left(t\right)-\sigma t$
appearing in the equation; (b) mitigation burden. }

\end{figure}

\begin{figure}
\includegraphics[scale=0.7]{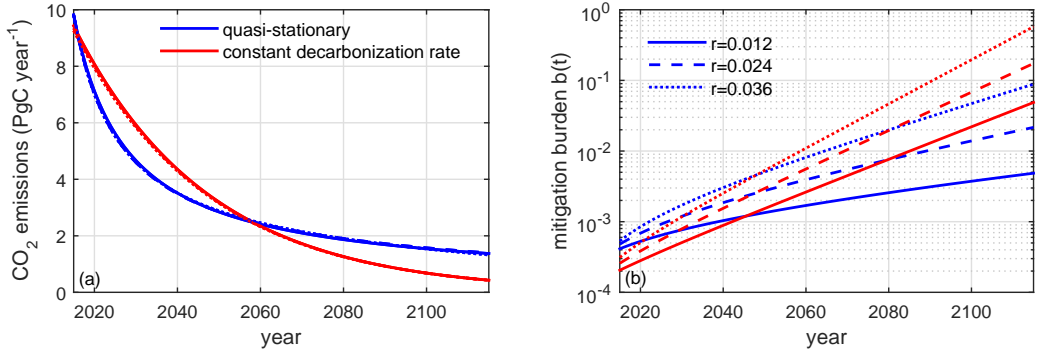}

\protect\caption{Effect of delay in decarbonization on future burden of meeting the
same cumulative emissions target of 300 PgC: (a) quasi-stationary
emissions pathway with early mitigation and a pathway with constant
decarbonization rate; (b) mitigation burden. Solid, dashed, and dotted
curves indicate GGDP growth rate of $0.012$, $0.024$, and $0.036$
respectively. The mitigation burden increases with time, and increases
more rapidly in scenarios where mitigation is postponed. A small fraction
($\leq$ $0.1$ \%) of GGDP is saved in the present time by choosing
the constant mitigation rate scenario, but leads to much higher mitigation
burden in the future (several percent of GGDP). }

\end{figure}

\subsection{Costs of mitigation}

Let us consider the overall costs of mitigation, by evaluating $f=\int_{0}^{T}e^{-\delta s}P\left(s\right)ds/\int_{0}^{T}e^{-\delta s}g\left(s\right)ds$,
the ratio between discounted expenditures and GGDP for the quasi-stationary
solution. Figure 7 shows that this is a convex function of cumulative
emissions goal, as Section 2.6 argued. The scale is logarithmic, and
mitigation cost increases rapidly for stringent cumulative emissions
goals as more rapid mitigation has diminishing effects on cumulative
emissions and furthermore expenditure rises. Still the costs are limited
to a small fraction of discounted GGDP, although this elides how the
burden of mitigation is distributed across time (Section 4.4). Background
assumptions about economic growth play an important role, and the
ratio in Figure 7 is smaller with higher discount rates since the
mitigation burden is increasing with time. Mitigation cost can vary
by more than an order of magnitude depending on the assumptions of
the model and the mitigation goal, but the results of our idealized
model seem to reproduce qualitatively the results from an integrated
assessment model (MESSAGE) discussed by \citet{Rogelj2015}. The graph
shows an approximately linear relationship on a logarithmic scale,
so the relation between costs $f$ and the cumulative emission goal
$M_{0}$ roughly obeys power law $f\left(M_{0}\right)=f_{1}\left(M_{0}/M_{01}\right)^{-n}$,
with $M_{01}=1000$ PgC being a reference goal and leading to cost
$f_{1}$. 

Figure 8a shows that higher economic growth increases the reference
cost $f_{1}$ substantially, and this is more sensitive to assumptions
about long-term economic growth than to the exponent in the MAC curve.
The power $n$ describes sensitivity $\frac{\triangle f/f}{-\triangle M_{0}/M_{0}}$
of relative changes in $f$ to relative changes in the cumulative
emissions goal. Figure 8b shows higher sensitivity to the cumulative
emissions goal in case the MAC curve rises sharply and GGDP rises
slowly. Rapid economic growth would make meeting cumulative emissions
goals more expensive regardless of how stringent they might be, whereas
sharply rising MAC makes cost more sensitive to cumulative emissions.
For median values the power in the above model $n\cong2.2$, so halving
the future cumulative emissions goal (for example from a $2$ C scenario
to a $1.5$ C scenario; recall that present warming is about $1.0$
C) would increase cost by a factor of about 4.6. 

\begin{figure}
\includegraphics[scale=0.7]{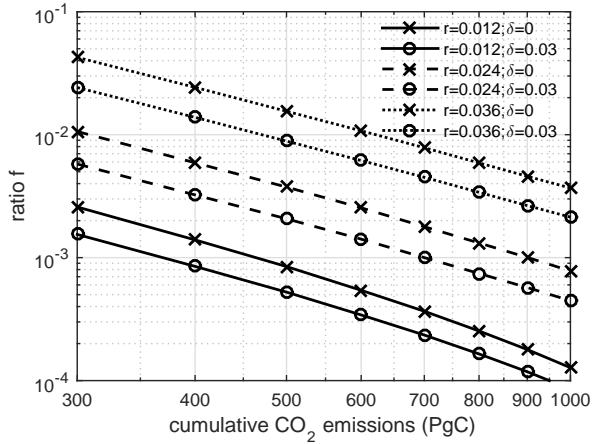}

\protect\caption{Ratio $f$ between discounted expenditures on mitigation and discounted
GGDP, graphed versus cumulative CO\protect\textsubscript{2} emissions.
Each curve shows results for a different combination of GGDP growth
rate and discount rate. For medium economic growth scenarios the expenditures
are limited to 1\% of discounted GGDP, but can be a few percent in
high growth scenarios. The ratio of expenditures and GGDP increases
with time (Figures 5 and 6), and therefore $f$ is smaller at higher
discount rates. The relationship is roughly linear on a logarithmic
scale, indicating a power law. }
\end{figure}

\begin{figure}
\includegraphics[scale=0.7]{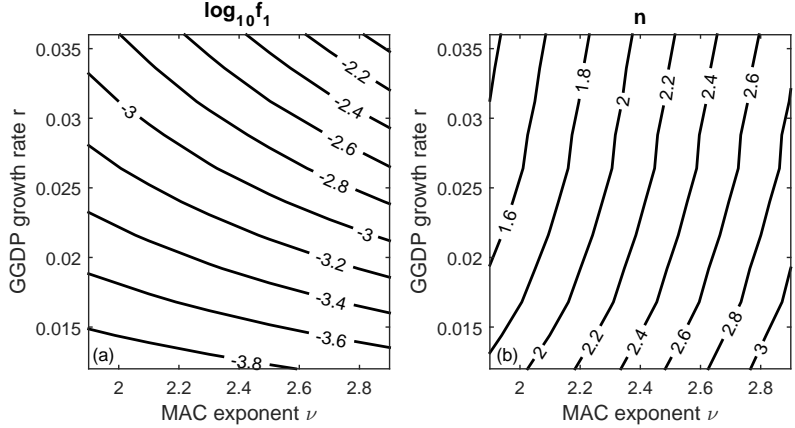}

\protect\caption{Parameters of power law between cumulative emissions goal and mitigation
cost, as a function of GGDP growth rate and exponent of MAC curve.
Mitigation cost is measured by $f$, the discounted expenditure as
fraction of discounted GGDP, and we fit the model: $f\left(M_{0}\right)=f_{1}\left(M_{0}/M_{01}\right)^{-n}$,
with $f_{1}$ being the cost of limiting $M_{0}$ to a reference value
of $M_{01}=1000$ PgC: (a) logarithm of reference cost $f_{1}$, which
is larger with higher growth rate and the MAC exponent, but is more
sensitive to the former; (b) exponent $n$ is higher with larger MAC
exponent and smaller economic growth. Mitigation cost is more sensitive
to cumulative emissions goal if MAC rises sharply and economic growth
is slow. Mitigation cost can vary more than an order of magnitude
depending on assumptions about the MAC and economic growth. }
\end{figure}

\section{Conclusions}

We have estimated the scale of expenditures involved in reducing cumulative
CO\textsubscript{2} emissions by reducing emissions intensity of
GGDP (\textquotedbl{}decarbonization\textquotedbl{}). Our model assumes
constant global income elasticity of CO\textsubscript{2} emissions,
which leads to exogenous decarbonization, independent of mitigation
policy, occurring in the business as usual scenario. This occurs at
rate increasing with the GGDP growth rate; and in case of constant
GGDP growth rate, the effects are analogous to discounting in time.
We also neglect decreases in marginal abatement costs with time, but
expect that real costs would have to decrease by a large degree to
undermine our main conclusions. Mitigation expenditure has two contributions,
arising from reducing emissions intensity and expanding effort as
the global economy grows. The first contribution is initially dominant
for the scenarios we have examined but the second contribution plays
a growing role. 

The MAC is estimated to rise steeply enough that the CO\textsubscript{2}
mitigation burden, defined as expenditure as a fraction of GGDP, is
expected to increase following equation (\ref{eq:18}). Future generations
would have to expend larger fractions of GGDP on decarbonization,
even in scenarios where the decarbonization rate is decreasing, as
cheaper mitigation options become exhausted. Rapid global economic
growth increases the mitigation burden on future generations, and
despite being wealthier in scenarios with higher-growth they would
have to spend larger fractions of GGDP as it becomes necessary to
ascend the MAC curve more rapidly. 

Discounted mitigation expenditures are convex functions of cumulative
CO\textsubscript{2} emissions. Over long time-horizons, larger rates
of decarbonization have diminishing benefits for reducing cumulative
CO\textsubscript{2} emissions, whereas mitigation expenditures increase
more rapidly. Global warming is approximately proportional to cumulative
emissions (\citet{Matthews2009,MacDougall2015,MacDougall2016,Tokarska2016}),
and the cost of emissions reduction is a convex function of the level
of global warming and expenditures increase more steeply if the policy
goal involves a smaller degree of global warming. Nonconvexities could
still arise in climate change economics due to non-convex damages
from, for example, threshold effects (\citet{Fisher1993,Lempert1996,Keller2004}). 

We measure overall costs by the discounted mitigation expenditures
as a fraction of discounted GGDP, and there is an approximate power
law relation between costs and cumulative emissions below 1000 PgC,
with exponent substantially larger than one. This exponent depends
mainly on the exponent in the MAC. Implications are best understood
through an example. Consider two alternate mitigation trajectories
where future cumulative emissions differ by a factor of two, for example
with global warming of 2.0 K versus 1.5 K where contributions from
other forcers are assumed to remain unchanged from present-day values.
With the exponent in the power law being generally larger than 2,
the more stringent trajectory would involve CO\textsubscript{2} mitigation
costs that are atleast 4 times larger. However such an account neglects
the benefits of early investments in low-carbon technologies and knowledge
spillovers to other sectors (\citet{Aghion2014,Dechezlepretre2014}). 

The optimization problem examined here seeks the pathway for global
decarbonization that minimizes discounted mitigation expenditures,
while meeting an exogenous constraint on cumulative CO\textsubscript{2}
emissions. Solving it required us to \textquotedbl{}regularize\textquotedbl{}
the variational problem because the original problem led to an algebraic
Euler-Lagrange equation whose solution did not satisfy the initial
condition on the integral of decarbonization rate. Regularization
yields a soluble Euler-Lagrange equation in the form of an initial
value problem in the integrated decarbonization rate. However the
solution carries economic meaning only for the special case of absence
of exogenous decarbonization and time-discounting. The general problem
of minimizing discounted mitigation expenditures in the presence of
exogenous decarbonization is important, but beyond our present scope.
Furthermore the solution is only \textquotedbl{}quasi-stationary\textquotedbl{},
with respect to a restricted class of perturbations leaving intact
the integrated decarbonization rate at the end of the time-horizon
as well. 

For the soluble case noted above, the optimal solution has decarbonization
rate proportional to emissions. An attraction of this solution is
that it does not depend on the parameters of the MAC model. However,
policy cannot be chosen without making long-term economic growth forecasts,
illustrating another difficulty of choosing climate policy under uncertainty
(\citet{Roughgarden1999,Pindyck2013}). The difficulty of forecasting
economic growth makes policy-choice in a cumulative CO\textsubscript{2}
framework uncertain. For example the levels of price or quantity instruments
are likely to determine the position on the MAC reached, but their
efficacy cannot be determined except ex-post once average growth rate
of GGDP during the long time-horizon of interest becomes known. 

The discounted mitigation expenditure is not only a function of cumulative
emissions, but depends also on the decarbonization pathway. Expenditures
are higher in scenarios requiring larger decarbonization rates in
the future during periods of much higher MAC. We illustrate using
the quasi-stationary solution for limiting cumulative emissions to
300 PgC over the next hundred years.%
\footnote{Although not being quasi-stationary in the presence of exogenous decarbonization,
it provides an archetype of a pathway involving early mitigation.%
} Although decarbonization starts rapidly at an initially higher rate,
the cost of mitigation amounts to a small burden at the present time
($<0.1\%$ of GGDP), but rising to 1\% or more at the end of the 100-year
period. For trajectories meeting this cumulative emissions constraint,
those with lower mitigation burdens in the present involve much higher
burdens in the future, and small savings in the present translate
to much larger future costs. These results are consistent with the
work of \citet{Vogt-Schilb2014} who suggest that near-term policy
should take into consideration longer-term targets as well, otherwise
costs of meeting the latter would be higher. 

Cumulative carbon accounting appears to make higher mitigation burdens
to future generations inevitable, but present choices can mitigate
some of this. While future expenditures can be discounted at a goods-discounting
rate based on the opportunity cost of capital (\citet{Nordhaus1993}),
discounting of future burdens can only be based on a positive value
of the pure rate of time-preference that reflects lower weights being
ceded to welfare of future generations.

\section*{Acknowledgments}

This work has been supported by Divecha Centre for Climate Change,
Indian Institute of Science. Thanks to several colleagues for helpful
suggestions.

\section*{Appendix 1: Derivation of quasi-stationary solution and degenerate
case}

For stationarity of $I\left(K,\dot{K}\right)=\int_{0}^{T}f\left(t,K,\dot{K}\right)dt+\lambda_{1}\left\{ \int_{0}^{T}m\left(t,K,\dot{K}\right)dt-M_{0}\right\} $
, we require the first-variation $\delta I$ in the integral due to
small changes $\delta K$ and $\delta\dot{K}$ to vanish. The variation
$\delta I=I\left(K+\delta K,\dot{K}+\delta\dot{K}\right)-I\left(K,\dot{K}\right)$
is 
\begin{equation}
\delta I=\int_{0}^{T}\left\{ \frac{\partial f}{\partial K}\delta K+\frac{\partial f}{\partial\dot{K}}\delta\dot{K}+\lambda_{1}\left(\frac{\partial m}{\partial K}\delta K+\frac{\partial m}{\partial\dot{K}}\delta\dot{K}\right)\right\} dt=0\label{eq:31}
\end{equation}
and, integrating by parts $\int_{0}^{T}\frac{\partial f}{\partial\dot{K}}\delta\dot{K}dt=\frac{\partial f}{\partial\dot{K}}\left(T\right)\delta K\left(T\right)-\int_{0}^{T}\frac{d}{dt}\left(\frac{\partial f}{\partial\dot{K}}\right)\delta Kdt$,
using $\delta K\left(0\right)=0$ and there is a corresponding equation
involving $m\left(t,K,\dot{K}\right)$. This yields
\begin{equation}
\delta I=\left\{ \frac{\partial f}{\partial\dot{K}}\left(T\right)+\frac{\partial m}{\partial\dot{K}}\left(T\right)\right\} \delta K\left(T\right)+\int_{0}^{T}\left\{ \frac{\partial f}{\partial K}-\frac{d}{dt}\left(\frac{\partial f}{\partial\dot{K}}\right)+\lambda_{1}\left(\frac{\partial m}{\partial K}-\frac{d}{dt}\left(\frac{\partial m}{\partial\dot{K}}\right)\right)\right\} \delta Kdt=0\label{eq:32}
\end{equation}
 for arbitrary changes $\delta K$. A subset of arbitrary changes
$\delta K$ involves those for which $\delta K\left(T\right)=0$ and
for this $\frac{\partial f}{\partial K}-\frac{d}{dt}\left(\frac{\partial f}{\partial\dot{K}}\right)+\lambda_{1}\left(\frac{\partial m}{\partial K}-\frac{d}{dt}\left(\frac{\partial m}{\partial\dot{K}}\right)\right)$
must vanish. This yields the Euler-Lagrange (E-L) equation (\ref{eq:21}),
which must also be satisfied when one of the endpoints, in this case
at $t=T$, is not fixed by the specification of the problem. In addition
the problem must satisfy the \textquotedbl{}natural boundary condition\textquotedbl{}
$\frac{\partial f}{\partial\dot{K}}\left(T\right)+\frac{\partial m}{\partial\dot{K}}\left(T\right)=0$
in order to be stationary for arbitrary changes $\delta K$. Readers
may refer to \citet{Brunt2004} for general discussion. 

However, is not possible to find a solution to our problem that satisfies
the aforementioned natural boundary condition. Therefore we consider
only changes involving $\delta K\left(T\right)=0$, so that for these
changes equation (\ref{eq:32}) is equivalent to the E-L equation.
The meaning of this is that, once the E-L equation with initial condition
$K\left(0\right)=0$ is solved for $K\left(t\right)$, this solution
is only stationary with respect to changes that preserve both $K\left(0\right)$
and $K\left(T\right)$. In this sense the solution is \textquotedbl{}quasi-stationary\textquotedbl{},
i.e. only relative to a restricted set of changes $\delta K$. 

Degeneracy arises in the following manner. In our problem, $f\left(t,K,\dot{K}\right)$
has form $\dot{K}f_{1}\left(t,K\right)+f_{2}\left(t,K\right)$ so
that $\frac{d}{dt}\left(\frac{\partial f}{\partial\dot{K}}\right)=\dot{K}\frac{\partial f_{1}}{\partial K}+\frac{\partial f_{1}}{\partial t}$
and $\frac{\partial f}{\partial K}=\dot{K}\frac{\partial f_{1}}{\partial K}+\frac{\partial f_{2}}{\partial K}$
, so that $\frac{\partial f}{\partial K}-\frac{d}{dt}\left(\frac{\partial f}{\partial\dot{K}}\right)=\frac{\partial f_{2}}{\partial K}-\frac{\partial f_{1}}{\partial t}$,
without any terms involving $\dot{K}$. The term $\frac{\partial m}{\partial K}-\frac{d}{dt}\left(\frac{\partial m}{\partial\dot{K}}\right)$
does not produce any terms involving $\dot{K}$ since $m=m\left(K,t\right)$,
and we are left with an algebraic E-L equation. Such degenerate cases
arise when dependence on $\dot{K}$ is linear (\citet{Brunt2004}).
To generate an initial value problem in $K$, regularization is necessary.
Therefore we introduce another contribution to the functional, involving
$h\left(t,K,\dot{K}\right)$, in Section 3.

\section*{Appendix 2: Solution to Euler-Lagrange equation for $\sigma>0$}

Consider equation (\ref{eq:29}) for $\delta=0$ but $\sigma>0$
\begin{equation}
\dot{x}\left(t\right)=\frac{\lambda_{1}\mu_{0}}{\lambda_{2}}n\left(t\right)-\frac{\sigma\beta}{\lambda_{2}}n\left(t\right)\left(x\left(t\right)\right)^{\nu}\label{eq:33}
\end{equation}
where $n\left(t\right)=e^{-\sigma t}g\left(t\right)$, and expanding
$x\left(t\right)\cong x_{0}\left(t\right)+\sigma x_{1}\left(t\right)$
in small parameter $\sigma\ll1$ and substituting in equation (\ref{eq:33})
\begin{equation}
\dot{x}_{0}\left(t\right)+\sigma\dot{x}_{1}\left(t\right)\cong\frac{\lambda_{1}\mu_{0}}{\lambda_{2}}n\left(t\right)-\frac{\sigma\beta}{\lambda_{2}}n\left(t\right)\left(x_{0}\left(t\right)\right)^{\nu}\left(1+\sigma\frac{x_{1}\left(t\right)}{x_{0}\left(t\right)}\right)^{\nu}\label{eq:34}
\end{equation}
with terms constant in $\sigma$ equating to $\dot{x}_{0}\left(t\right)=\frac{\lambda_{1}\mu_{0}}{\lambda_{2}}n\left(t\right)$
with $x_{0}\left(0\right)=1$ and those of first-degree in $\sigma$
yielding 
\begin{equation}
\dot{x}_{1}\left(t\right)=-\frac{\beta}{\lambda_{2}}n\left(t\right)\left(x_{0}\left(t\right)\right)^{\nu}\label{eq:35}
\end{equation}
with $x_{1}\left(0\right)=0$. The $0$\textsuperscript{th}- degree
equation is integrated for $x_{0}\left(t\right)=1+\frac{\lambda_{1}\mu_{0}}{\lambda_{2}}N\left(t\right)$,
where $N\left(t\right)=\int_{0}^{t}n\left(s\right)ds$. 

Then $x_{1}\left(t\right)=-\int_{0}^{t}\frac{\beta}{\lambda_{2}}n\left(s\right)\left(1+\frac{\lambda_{1}\mu_{0}}{\lambda_{2}}N\left(s\right)\right)^{\nu}ds$
, which simplifies to $x_{1}\left(t\right)=-\frac{\beta}{\left(\nu+1\right)\lambda_{1}\mu_{0}}\left\{ \left(1+\frac{\lambda_{1}\mu_{0}}{\lambda_{2}}N\left(t\right)\right)^{\nu+1}-1\right\} $.
With $\frac{\lambda_{1}\mu_{0}}{\lambda_{2}}N\left(t\right)\ll1$
we can further approximate for $x_{1}\left(t\right)=-\frac{\beta}{\lambda_{2}}N\left(t\right)$
and using $\beta=\alpha\mu_{0}$ 
\begin{equation}
x\left(t\right)\cong1+\frac{\mu_{0}}{\lambda_{2}}\left(\lambda_{1}-\sigma\alpha\right)N\left(t\right)\label{eq:36}
\end{equation}
As it turns out, $\lambda_{1}$ is small compared to $\sigma\alpha$
so that $\lambda_{1}-\sigma\alpha<0$ in general, and $x\left(t\right)$
is decreasing in time for $\sigma>0$ . Hence the Euler-Lagrange solution
for $\sigma>0$ does not correspond to relevant economic optima, and
such a situation cannot be solved using the method of regularization
described here. Analogously it can be shown that with $\delta>0$
the solution $x\left(t\right)$ and hence $K\left(t\right)$ is decreasing
in time. Therefore we solve for the quasi-stationary pathway following
equation (\ref{eq:30}), but with $\sigma$ present in the emissions
model and affecting how the cumulative emissions goal is met. Once
this pathway is estimated, it is applied to our model of mitigation
expenditures having in general nonzero $\sigma$ and $\delta$. 

\bibliographystyle{27C__Dropbox_SubmittedPapers_0_OptimalMitigation_agufull08}
\bibliography{26C__Dropbox_SubmittedPapers_0_OptimalMitigation_Optimal_CO2_mitigation_refs}

\end{document}